\documentclass[referee,sn-nature]{sn-jnl}

\usepackage{graphicx}%
\usepackage{multirow}%
\usepackage{amsmath,amssymb,amsfonts}%
\usepackage{amsthm}%
\usepackage{mathrsfs}%
\usepackage[title]{appendix}%
\usepackage{xcolor}%
\usepackage{textcomp}%
\usepackage{manyfoot}%
\usepackage{booktabs}%
\usepackage{algorithm}%
\usepackage{algorithmicx}%
\usepackage{algpseudocode}%
\usepackage{listings}%

\theoremstyle{thmstyleone}%
\theoremstyle{thmstyletwo}%

\theoremstyle{thmstylethree}%

\begin{document}

\title[Programmable Photonic Simulator for Spin Glass Models]{Programmable Photonic Simulator for Spin Glass Models}


\author*[1]{\fnm{Weiru} \sur{Fan}}\email{weiru\_fan@zju.edu.cn}
\equalcont{These authors contributed equally to this work.}

\author[1]{\fnm{Yuxuan} \sur{Sun}}\email{yuxuan\_sun@zju.edu.cn}
\equalcont{These authors contributed equally to this work.}

\author[1]{\fnm{Xingqi} \sur{Xu}}\email{xuxingqi@zju.edu.cn}

\author*[1,2,3,4]{\fnm{Da-Wei} \sur{Wang}}\email{dwwang@zju.edu.cn}

\author[1,2,3]{\fnm{Shi-Yao} \sur{Zhu}}\email{syzhu@zju.edu.cn}

\author[1]{\fnm{Hai-Qing} \sur{Lin}}\email{hqlin@zju.edu.cn}

\affil[1]{\orgdiv{Zhejiang Province Key Laboratory of Quantum Technology and Device, School of Physics, and State Key Laboratory for Extreme Photonics and Instrumentation}, \orgname{Zhejiang University}, \orgaddress{\city{Hangzhou}, \postcode{310027}, \state{Zhejiang Province}, \country{China}}}

\affil[2]{\orgdiv{College of Optical Science and Engineering}, \orgname{Zhejiang University}, \orgaddress{\city{Hangzhou}, \postcode{310027}, \state{Zhejiang Province}, \country{China}}}

\affil[3]{\orgname{Hefei National Laboratory}, \orgaddress{\city{Hefei}, \postcode{230088}, \country{China}}}

\affil[4]{\orgdiv{CAS Center for Excellence in Topological Quantum Computation}, \orgname{University of Chinese Academy of Sciences}, \orgaddress{\city{Beijing}, \postcode{100190}, \country{China}}}


\abstract{Spin glasses  featured by frustrated interactions and metastable states have important applications in chemistry, material sciences and artificial neural networks. However, the solution of the spin glass models is hindered by the computational complexity that exponentially increases with the sample size. Photonic Ising machines based on spatial light modulation can speed up the calculation by obtaining the Hamiltonian from the modulated light intensity. However, the large-scale generalization to various spin couplings and higher dimensions is still elusive. Here, we develop a Fourier-mask method to program the spin couplings in photonic Ising machines. We observe the phase transition of the two-dimensional Mattis model and the J$\mathrm{_1}$-J$\mathrm{_2}$ model and study the critical phenomena. We also demonstrate that the three-dimensional Ising model, which has not been analytically solved, can be effectively constructed and simulated in two-dimensional lattices with Fourier masks. Our strategy provides a flexible route to tuning couplings and dimensions of statistical spin models, and improves the applicability of optical simulation in neural networks and combinatorial optimization problems.}

\maketitle

\section{Introduction}
The spin glass model (SGM) \cite{edwards1975theory} provides a unique perspective to understand various disordered systems with complex interactions across multiple disciplines, including brain science \cite{amit1989modeling}, quantum chromodynamics \cite{halasz1998phase}, and network topologies \cite{herrero2002ising}. In particular, the SGM provides intriguing tools for probability graphs in machine learning \cite{fan2023searching,eaton2012spin} and combinatorial optimization problems \cite{barahona1988application,stein2013spin,yamashita2023low} such as protein folding \cite{bryngelson1987spin}. However, the calculation of large-scale SGM is challenging on conventional computers due to the exponentially increasing configuration space for a large number of spins. To meet this challenge, the SGM has been tackled by analogue computation in various physical systems \cite{sarkar2005synthesizing,johnson2011quantum,marandi2014network,honjo2021100,inaba2022potts,pierangeli2019large,fang2021experimental,huang2021antiferromagnetic,jacucci2022tunable,leonetti2021optical}. In particular, the photonic Ising machine (PIM), which encodes the spins on the wavefront of light, is an efficient platform for simulating the large-scale SGM  by accelerating part of the computing task with photonic processors \cite{wetzstein2020inference,zhou2022photonic}. \par 

The programmability of the spin couplings is a crucial requirement for a universal PIM to simulate the phase transitions of various SGMs \cite{fisher1988equilibrium,temesvari2010ising,baity2018aging}. Such a programmable PIM can also be used to solve non-deterministic polynomial time (NP)-hard problems by mapping combinatorial optimization to finding the ground states of SGMs with specific spin couplings \cite{wu1982potts,lucas2014ising,mohseni2022ising}. However, most of the existing PIMs are restricted to models with fixed spin couplings, limiting their practical applications. Recently, wavelength-division multiplexing \cite{luo2023wavelength}, time multiplexing \cite{yamashita2023low} and phase-encoding of eigen-decomposed Hamiltonian \cite{ouyang2022demand} are proposed to synthesize arbitrary spin couplings in PIM. However, these strategies consume excessive spatial or temporal resources of the spatial phase modulator (SLM) to achieve tunability in couplings, and thus far have been limited to small-scale simulation. \par 

In this Letter, we propose and implement a programmable Fourier-mask method to extract energy contributions of different spin pairs from the all-to-all couplings on the Fourier domain, which allows us to solve various large-scale SGMs. The feasibility of this method is experimentally verified with the 2D Mattis model with nearest-neighbor (NN) couplings and the $\mathrm{J_1}$-$\mathrm{J_2}$ SGMs with both NN and next-nearest-neighbor (NNN) couplings. We observe the phase transition and determine the critical parameters in each model. We further use the Fourier-mask PIM to solve the 3D Ising model on a simple cubic lattice. Such a programmable PIM has promising applications in solving large-scale NP-hard problems in data science, biology, and sociology.

\section{Results}
\begin{figure}[t]
\includegraphics[width=0.85\linewidth]{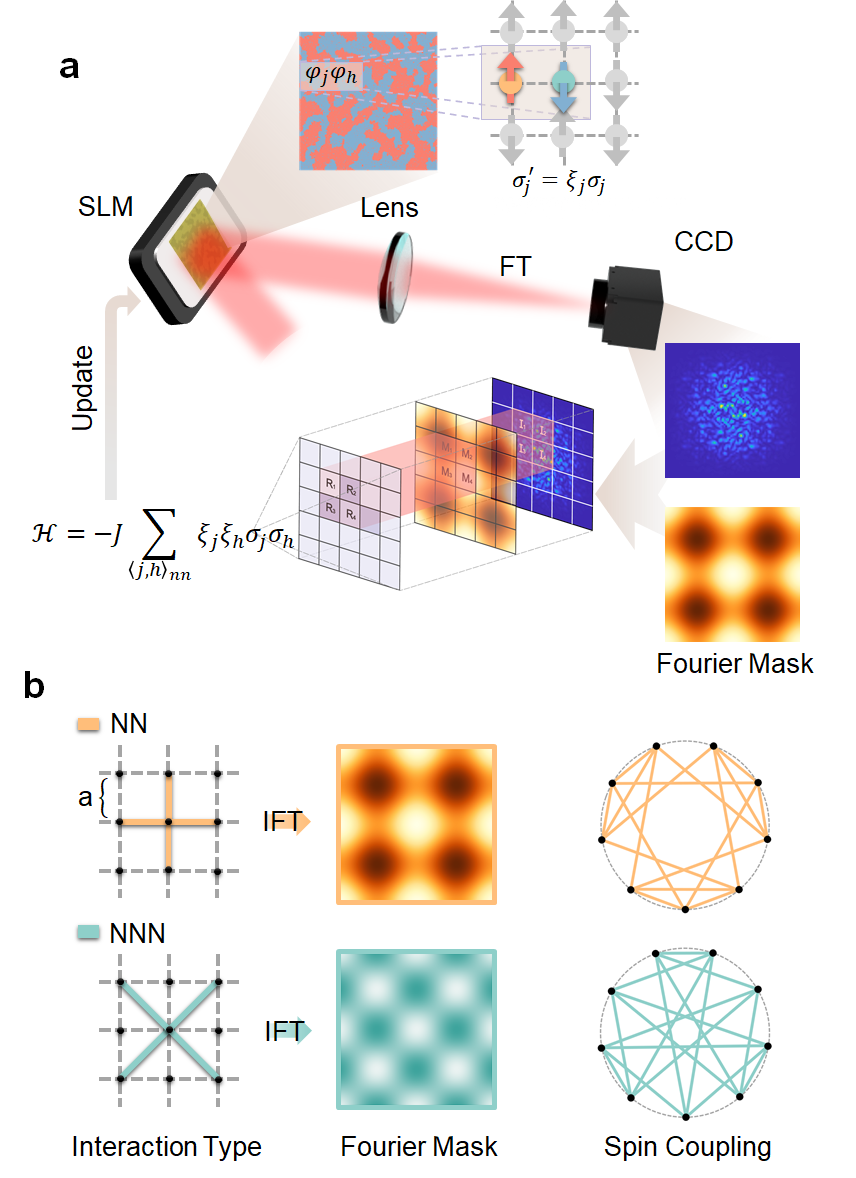}
\caption{\label{fig_1} The principle of Fourier-mask PIMs. (a) The configuration of the PIM with a Fourier mask. (b) The Fourier masks with NN and NNN couplings. In our scheme, the effective spin values $\sigma_j^{\prime}=\xi_j\sigma_j$ are encoded as $\varphi_j$ on the light front by a phase-type SLM. The modulated light passes through a lens to perform the Fourier transform (FT), and is recoded by a charge-coupled device (CCD) in the Fourier plane. The Hamiltonian of model is evaluated by the element-wise product of the field intensity and the Fourier mask. To update spin configuration, classical annealing algorithm is used to evolve towards the ground state. Inverse Fourier transform (IFT) is used in designing the Fourier mask. In the spin coupling graph, the dots denote the spins and the solid lines are the couplings between them.}
\end{figure}

\subsection{Fourier-mask PIM}
The Hamiltonian of the Mattis SGM is \cite{mattis1976solvable},
\begin{eqnarray}
\mathcal{H} = -\sum_{[jh]}J_{jh}\sigma_j\sigma_h =-J\sum_{[jh]}\xi_j\xi_h\sigma_j\sigma_h,
\label{equ_1}
\end{eqnarray}
where $J_{jh} = J\xi_j\xi_h$ is the coupling strength between the spins on sites $j$ and $h$, $J$ is the interaction strength constant. The random variables $\xi_j=2X-1$ where $X$ obeys binary distribution $B(1,p)$ with $p$ being the probability for $X=1$. The values of the spin $\sigma_j = 1$ or $-1$. The  symbol $[jh]$ in the summation denotes the specific spin pairs such as $\langle jh\rangle$ for NN and $\langle\langle jh\rangle\rangle$ for NNN couplings. In conventional photonic simulation, the amplitude and phase of light are used to simulate $\xi_j$ and $\sigma_j$. Both parameters must be controlled independently with positive and negative values. Here we adopt a gauge transformation method \cite{fang2021experimental} to achieve the simulation on a single modulator, where each effective spin $\sigma_j^{\prime}=\xi_j\sigma_j$ can be encoded on an individual pixel (see Supplemental Materials for detail). The independent spin sites are coupled by a lens to perform optical Fourier transform. The intensity at the center of the Fourier plane has been used to solve the all-to-all coupling Ising model \cite{pierangeli2019large,fang2021experimental}. It has also been shown that antiferromagnetic couplings can be obtained by integrating the light intensity with proper functions \cite{huang2021antiferromagnetic}. The central task of this Letter is to utilize the the intensity distribution $I(\boldsymbol{x})$ in the Fourier plane to obtain the interaction energies of different types of spin couplings. \par 

The key idea of our approach in obtaining the desired Hamiltonian is to put a Fourier mask $I_M(\boldsymbol{x})$ on the light intensity  $I(\boldsymbol{x})$ (the size effect of pixels on the SLM is neglected),
\begin{eqnarray}
\int I(\boldsymbol{x}) I_M(\boldsymbol{x}) d \boldsymbol{x}=I_0 \sum_{j\neq h=1}^N C_{j h} \xi_j \xi_h \sigma_j \sigma_h+cI_0N,
\label{equ_2}
\end{eqnarray}
where the coefficients $C_{j h}$ are related to $I_M$ through a Fourier transform, $C_{j j}\equiv c$ is a constant obtained from $I_M$, and $I_0$ is a factor characterizing the overall intensity (see Supplemental Materials). We use Eq.~(\ref{equ_2}) to obtain various Mattis SGM Hamiltonians by choosing different Fourier masks $I_M$ according to the interaction types and spin couplings. The computation of Eq.~(\ref{equ_2}) can be carried out by directly imposing a photonic mask on the Fourier plane. 
We obtain $I_M$ by making the inverse Fourier transform of $C_{j h}$. The all-to-all interaction is generated by setting a uniform $C_{j h}=1$, which corresponds to a Dirac delta function of $I_M$. Therefore, the intensity at the center of the Fourier plane is used to calculate the all-to-all coupled Ising Hamiltonian \cite{fang2021experimental}. In general, the Fourier mask $I_M$ can be calculated by
\begin{eqnarray}
I_M(\boldsymbol{x})=\frac{1}{2 \pi} \int C_{j 1} \exp \left(-i \frac{2 \pi}{f \lambda} \boldsymbol{u}_j  \cdot\boldsymbol{x}\right) 
  d\boldsymbol{u}_j,
\label{equ_3}
\end{eqnarray}
where $f$ is the focal length of the lens, $\lambda$ is the wavelength of light, $\boldsymbol{u}_j$ is the position of the $j$th pixel on the SLM and we set $\boldsymbol{u}_1=0$.\par 

The Hamiltonians of different PIMs can be obtained by designing the corresponding Fourier masks (Fig. \ref{fig_1}a). For instance, the NN couplings on a square lattice can be synthesized by the inverse Fourier transform of the summation of four Dirac delta functions (Fig. \ref{fig_1}b). Mathematically, the Fourier mask is the summation of two cosine functions in $x$ and $y$ directions, $\mathrm{cos}(bx) +\mathrm{cos}(by)$, where $b$ is a coefficient determined by the system parameters. For the NNN couplings, the Fourier mask is designed with the same procedure to be $\mathrm{cos}(bx)\cdot\mathrm{cos}(by)$ (Fig. \ref{fig_1}b). These functions are shifted and renormalized to values bounded by 0 and 1 for experimental implementation with an intensity antenuator (see Methods and Supplemental Materials).\par 

A key problem in using Eq.~(\ref{equ_2}) to calculate the Hamiltonian in Eq. (\ref{equ_1}) is to eliminate the intensity factor $I_0$ to maintain consistent results with different illuminating powers. The conventional wisdom is to set all spins in the same direction to produce an Airy disk, with its position being defined as the coordinate origin and the maximum intensity being set as the normalization factor \cite{huang2021antiferromagnetic}. However, the Airy spot has a finite size, resulting in deviations of the origin and the intensity factor $I_0$ from their actual values, and consequently a multiplicative bias in evaluating the Hamiltonian. Here, we develop an unbiased method with a weighting factor, which is the ratio between the Hamiltonian of the all aligned spin configuration and the integration in Eq.~(\ref{equ_2}) for the same configuration. Normalized with this weighting factor, the multiplicative bias is eliminated and the problem of defining the absolute illuminating power is circumvented (see Methods and Supplemental Materials). Such an accurate evaluation of the Hamiltonian lays the basis for implementing Monte Carlo algorithm with Metropolis-Hastings acceptance rule on the PIM to update the spin configurations, such that we can generate a reliable Markov chain and investigate the phase transition.\par 

\subsection{Mattis model with NN couplings}
To show the essence of our approach, we first simulate the Mattis SGM with the NN couplings. In this model, the random variables $\xi_j$ take the values 1 and $-1$ with the probabilities $p$ and $1-p$. It has three stable phases determined by the temperature $T$ and probability $p$. These three phases are characterized by two order parameters, the average magnetization $M$ and the spin glass order parameter $Q$, which can be defined as  $M\left(T, p\right)=	[ \left \langle \sum_i\sigma_i \right \rangle_T /N ]_p$, $Q\left(T, p\right)=	[ \left \langle \sum_i\sigma_i \right \rangle_T^2 /N^2 ]_p$, where $\left \langle \  \right \rangle_T$ is the average of samples on the Markov chain with fixed temperature $T$ and configuration $\left\{ \xi_j \right\}$, and $\left[ \  \right]_p$ is the average over different configurations $\left\{ \xi_j \right\}$ with the same probability $p$ \cite{nishimori2001statistical}. In the parameter space of $M$ and $Q$, the three phases are paramagnetic (PM) phase for $M=0$, $Q=0$, ferromagnetic (FM) phase for $M>0$, $Q>0$, and Mattis spin glass (MSG) phase for $M=0$, $Q>0$.\par

\begin{figure}[t]
\includegraphics[width=0.8\linewidth]{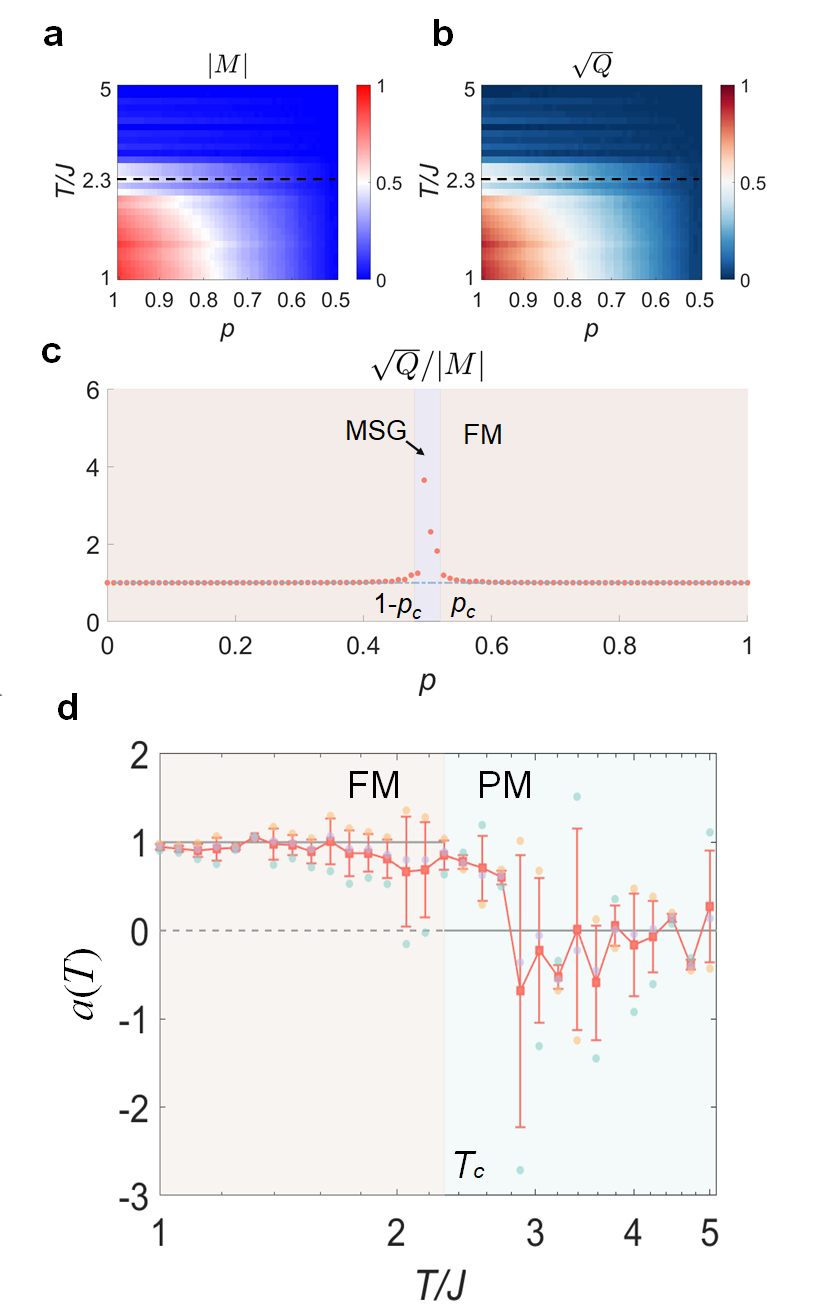}
\caption{\label{fig_2} Photonic simulation of the phase transition of the Mattis SGM with NN couplings. (a) $\left|M\right|$ and (b) $\sqrt{Q}$ as functions of $T$ and $p$. The transition from the PM to FM phase occurs at the critical temperature $T_c=2.27J$ (black dashed line). (c) $\sqrt{Q}/\left|M\right|$ as a function of $p$ at $T=J$. The value diverges for $1-p_c<p< p_c$ with a critical probability $p_c=0.52$. The blue dash-dotted line indicates $\sqrt{Q}/\left|M\right|=1$. (d) The size-scaling exponents. The orange, cyan and purple circles are $a(T)$ obtained from the total spin number combinations $(N1, N2) = (400, 100)$, $(400, 900)$ and $(900, 100)$. The red squares are the average values and the error bars indicate the standard deviation of three independent groups of data. The gray solid line is $a(T)$ predicted by the scaling theory with the critical temperature $T_c = 2.27J$.}
\end{figure}

The order parameters $M$ and $Q$ are computed with the Fourier-mask PIM at different $T$ and $p$ with $N=900$ spins (Fig. \ref{fig_2}a and \ref{fig_2}b). When $p=1$, the system is a 2D Ising model. The phase transition from PM to FM occurs around $T=2.3J$, consistent with Onsager’s exact solution $T_c\approx2.27J$ \cite{onsager1944crystal}. Below $T_c$, the spins are in the FM phase, and $\left|M\right|\approx 1$. Near $T_c$ the magnetization $\left|M\right|$ gradually changes from 1 to 0 due to the critical slowing down of the Monte Carlo algorithm \cite{binder1986spin}. For $T>T_c$ the spins are in the PM phase and $\left|M\right|\approx 0$. For a temperature $T<T_c$, an MSG phase for $1-p_c<p< p_c$ (with $p_c$ being the critical probability) emerges with $M=0$ but $Q>0$, i.e., characterized by a finite magnetization in random directions  \cite{fang2021experimental}. In Fig. \ref{fig_2}c, the phase transition between the FM and MSG phase is evident from the diverging values of $\sqrt{Q}/\left|M\right|$ at a low temperature $T=J$. The observed $p_c$ is consistent with the mean-field prediction $p_c=0.5167$ \cite{binder1986spin}.\par 

The size-scaling properties of the physical observables are important for thermal-dynamic-limit PIM. According to the scaling theory, the susceptibility per spin is $\chi=N(\left \langle M^2 \right \rangle_T-\left \langle M \right \rangle_T^2)/k_BT$, where $N\left \langle M^2 \right \rangle_T\propto N^{a(T)}$ with $a(T)=1$ for $T\textless T_c$ and $a(T)=0$ for $T\textgreater T_c$, and $k_B$ being the Boltzmann constant (we set $k_B=1$) \cite{miyashita1978monte}. To verify such scaling, two systems with different sizes $(N_1, N_2)$ are used to calculate the parameter $a(T)$ according to the relationship $a(T)=\mathrm{ln}(\left \langle M^2 \right \rangle_{T,N_1}/\left \langle M^2 \right \rangle_{T,N_2})/\mathrm{ln}\left(N_1/N_2\right)+1$, where $\left \langle M^2 \right \rangle_T$ is calculated with $N=100$, 400, and 900 (Fig. \ref{fig_2}d). We observe that below and above $T_c$ the average values of $a(T)$ are around 1 and 0, respectively. The exponent $a(T)$ has a larger variance in PM phase than in FM phase, because $\left \langle M^2 \right \rangle_T$ measures the fluctuation of magnetization above $T_c$. \par 

\begin{figure}[t]
\includegraphics[width=1\linewidth]{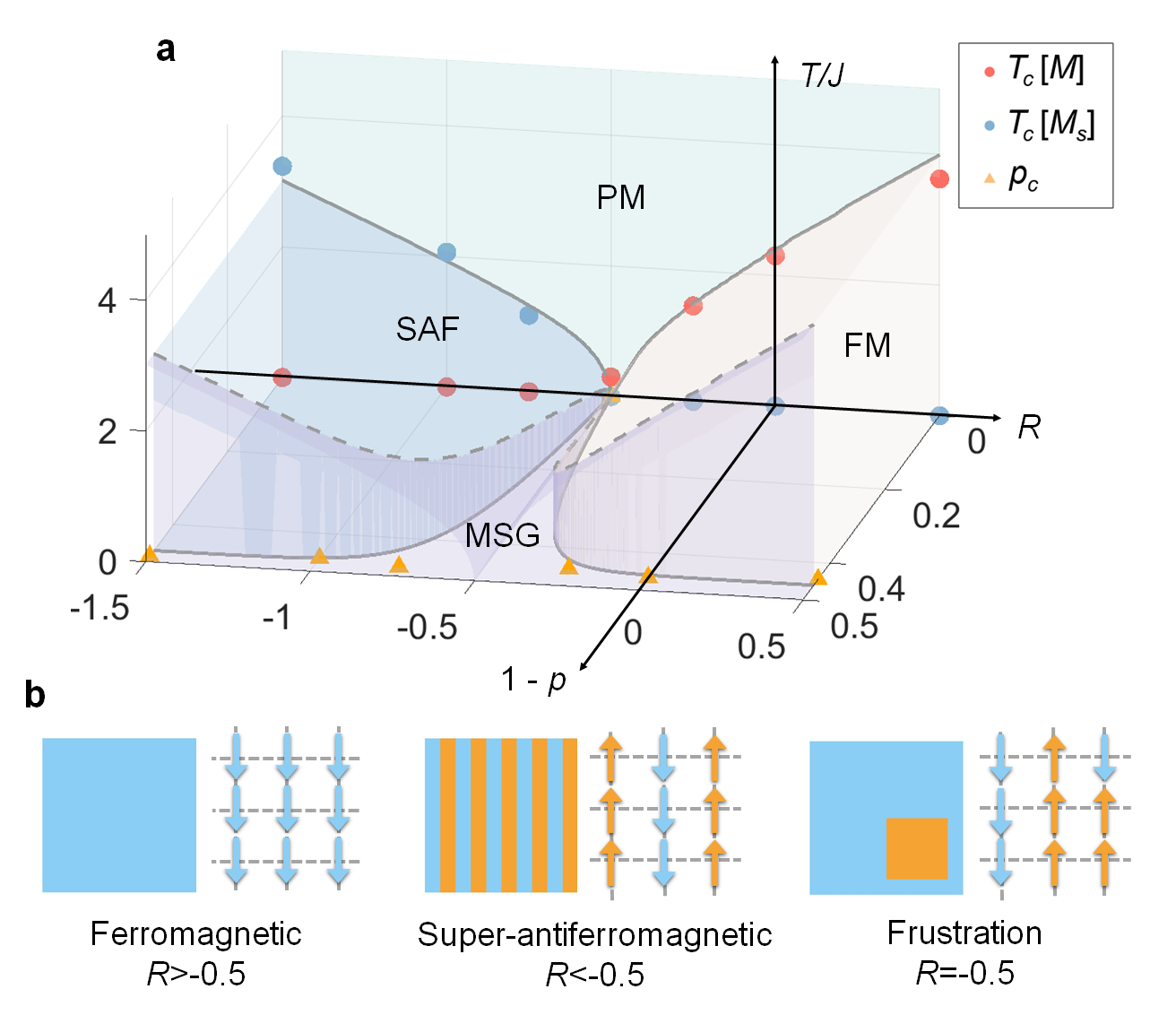}
\caption{\label{fig_3} The Fourier-mask PIM of the $\mathrm{J_1}$-$\mathrm{J_2}$ model. (a) The phase diagram of the $\mathrm{J_1}$-$\mathrm{J_2}$ model. (b) Ground state spin configurations for different phases. The gray solid lines in the $T$-$R$ plane and the $R$-$p$ plane are the phase boundaries from the group renormalization method \cite{nauenberg1974critical} and Monte Carlo simulation, respectively. The circles and triangles in (a) are the $T_c$’s and $p_c$’s obtained from the Fourier-mask PIM. The total number of spin is $N=100$.}
\end{figure}

\subsection{J$_1$-J$_2$ model}
The Fourier-mask PIM is programmable to synthesize SGMs with multiple types of couplings by combining the corresponding Fourier masks. We demonstrate such flexibility with the $\mathrm{J_1}$-$\mathrm{J_2}$ model, which has competing NN and NNN couplings with the Hamiltonian,
\begin{eqnarray}
\mathcal{H} = -J_1\sum_{\left \langle jh \right \rangle}\xi_j\xi_h\sigma_j\sigma_h -J_2\sum_{\left\langle\left \langle jh \right \rangle\right\rangle}\xi_j\xi_h\sigma_j\sigma_h,
\label{equ_3}
\end{eqnarray}
where $J_1$ and $J_2$ are the amplitudes of the NN and NNN couplings. We assume $R=J_2/J_1$ and $J_1=J>0$ for convenience. When $R<0$, the NN spin pairs prefer the ferromagnetic configuration, whereas the NNN pairs prefer the antiferromagnetic configuration. Such competition results in the emergence of a new phase \cite{xu2019higher} characterized by the staggered magnetic moment $M_s=\left(M_1-M_2\right)/2$, which is an order parameter for antiferromagnetic long-range order with $M_1$ and $M_2$ being the average magnetic moments of the two sublattices (e.g., even and odd rows or columns).\par 

We simulate the $\mathrm{J_1}$-$\mathrm{J_2}$ model in the Fourier-mask PIM with different $R$’s (see Fig. \ref{fig_3}). For $p=1$, $T_c$ is determined by the abrupt changes in $\left|M\right|$ or $\left|M_s\right|$ (Fig. \ref{fig_3}a). From the calculated values of $\left|M\right|$ and $\left|M_s\right|$, we divide the $T$-$R$ phase diagram into three regions, the PM, FM and super-antiferromagnetic (SAF) phases. In the $R$-axis, when $T=0$ and $p=1$, the ground state energy has a transition between SAF and FM phase at $R=-0.5$ \cite{nauenberg1974critical}. In general, when $R>-0.5$, the model is dominated by the ferromagnetic NN coupling, resulting in nonzero $M$ and zero $M_s$ below $T_c$, indicating the FM phase (Fig. \ref{fig_3}b left). When $R<-0.5$ the model is dominated by the anti-ferromagnetic NNN coupling, resulting in zero $M$ and nonzero $M_s$, and the ground state is characterized by a stripe SAF phase (Fig. \ref{fig_3}b center).\par

For $R=-0.5$, the NN and the NNN couplings are balanced, which results in the absence of long-range order (FM or SAF) at any finite temperature \cite{yin2009phase,selke1980two} (the corresponding ground state is shown in Fig. \ref{fig_3}b right). For different $R$’s, we calculate $\sqrt{Q}/\left|M\right|$ or $\sqrt{Q}/\left|M_s\right|$ to obtain the critical probability $p_c$ from the divergent points. When $R$ approaches to $-0.5$, the long-range order is increasingly prone to be destroyed by the disorder in the couplings such that the critical probability $p_c$ tends to 1, indicating that the point $(R, T, p) = (-0.5, 0, 1)$ is a critical quadruple point, which agrees with the theoretical predictions.\par 

\begin{figure}[t]
\includegraphics[width=1\linewidth]{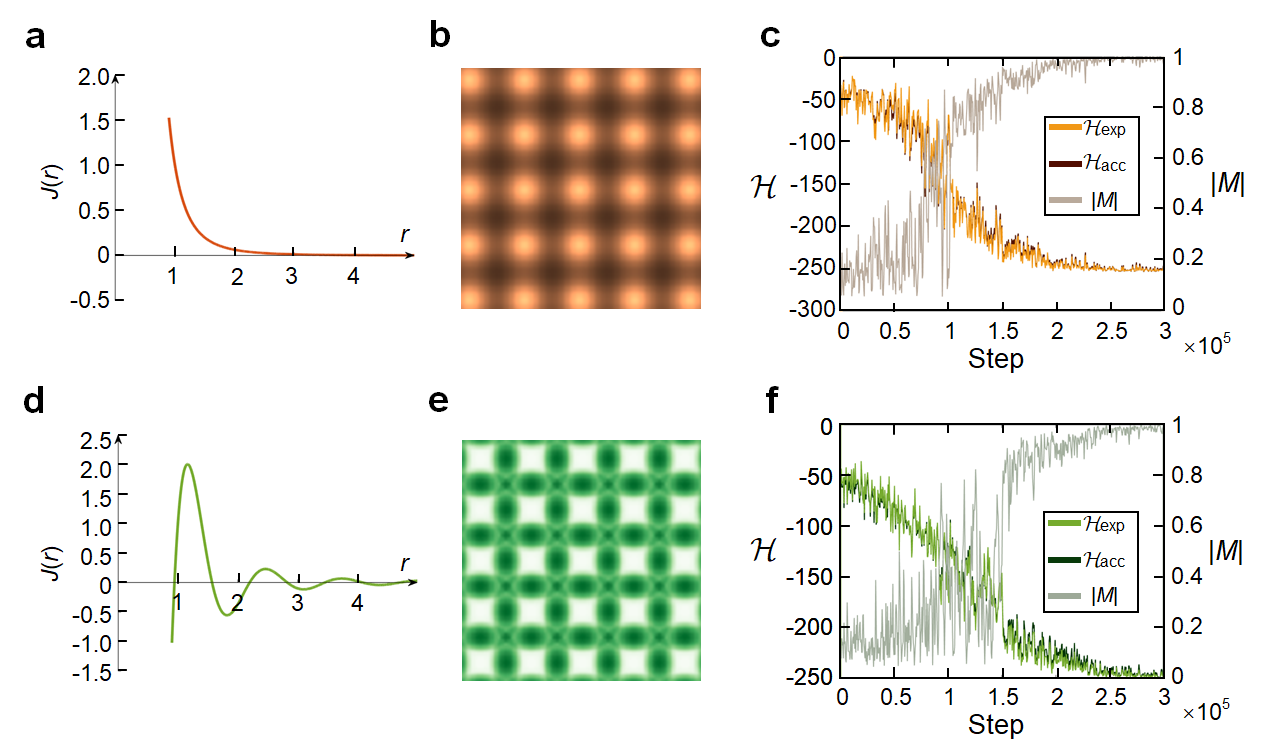}
\caption{\label{fig_4} {Fourier-mask photonic simulation of the annealing of SGMs with long-range interactions. (a) The interaction function, (b) the Fourier mask and (c) the Hamiltonian and magnetization as a function of the Monte Carlo steps during the annealing for a power-law decaying interaction $J(r)=1/r^4$. The distance $r$ is in the unit of lattice constant. (d)-(f) are those for the RKKY long-range interaction $J(r)=\cos(2k_F r)/r^3$ with $k_F=2.5$. In (c) and (f), the interaction Hamiltonian $\mathcal{H}_{\text{exp}}$ and magnetization $|M|$ are obtained from the photonic simulation, in comparison to $\mathcal{H}_{\text{acc}}$ from Monte Carlo simulation on an electronic computer. {During the annealing the temperature exponentially decreases from $7J$ to $J$ in 20 steps (equally distributed in the exponential variables). In each step the spins are flipped 15,000 times. Each data point is the averaged value from 500 spin configurations with the same spin couplings.} The total number of spins is $N=100$.}}
\end{figure}

The $\mathrm{J_1}$-$\mathrm{J_2}$ model can be viewed as the simplest SGM with long-range interactions. In traditional methods, the long-range interaction substantially increases the computational complexity~\cite{muller2023fast}, since all $N(N-1)/2$ spin pairs need to be considered in calculating the Hamiltonian. However, the range of interaction brings no difference for PIM. We verify the efficiency of Fourier-mask photonic simulation in the annealing of such SGMs (see Fig.~\ref{fig_4} (a)-(c) for the power-law decaying interaction~\cite{Chri2020Aging} and Fig.~\ref{fig_4} (d)-(f) for the Ruderman-Kittel-Kasuya-Yoshida (RKKY) interaction~\cite{Priour2004Disordered}). The Hamiltonian and magnetization are efficiently and accurately calculated by the corresponding PIM, demonstrating its effectiveness in simulating large-scale SGMs with long-range interactions.\par

\begin{figure}[t]
\includegraphics[width=0.9\linewidth]{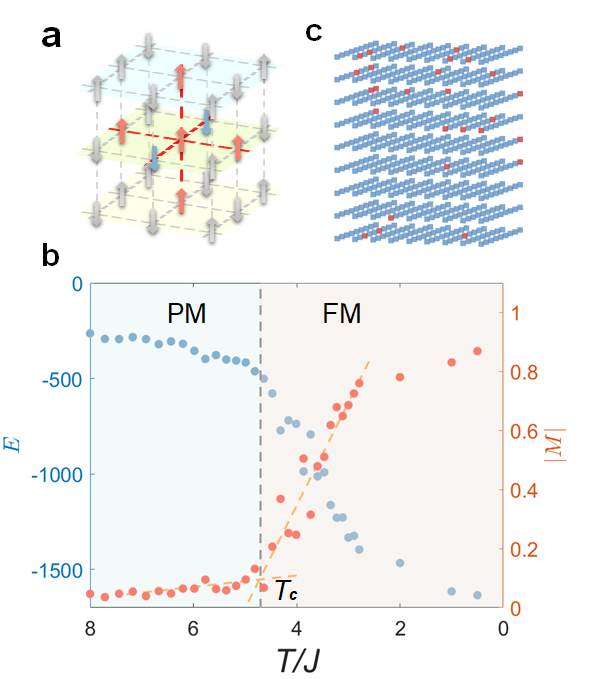}
\caption{\label{fig_5} Photonic simulation of the 3D Ising model. (a) Ising model on a 3D simple cubic lattice with NN coupling. (b) $E$ and the $\left|M\right|$ as functions of the temperature. (c) The ground state configuration of the 3D Ising model with NN couplings for $T=0.5J$. The two dashed lines in (b) are obtained by linearly fitting the data with $T$ and $\left|M\right|$ in the two phases. We use their cross point to determine $T_c =4.7J$. The number of spins $N=729$. The bule and red squares in (c) represent the two states of spins.}
\end{figure}

\subsection{3D Ising model}
The 3D Ising model has no known analytical solutions even for the simple cubic lattices (Fig. \ref{fig_5}a). To obtain the critical parameters, numerical methods such as Monte Carlo simulation \cite{preis2009gpu} and transfer matrix \cite{ghaemi2001calculation} have been used and a ferromagnetic phase transition was found at $T_c\approx4.5J$. The Fourier-mask can map the high-dimensional lattices into 2D lattices and thus can provide an efficient way to simulate high-dimensional Ising models. We simulate a 3D Ising model in a plane of 3×3 array and 9×9 spins in each array element, in total 729 spins (see Fig. \ref{fig_5}a). We design a Fourier mask to simultaneously achieve the intra-layer and inter-layer NN spin couplings (see Supplemental Materials for detail).\par 

Similar to the 2D Ising model, the order parameter $\left|M\right|$ is used to define the critical temperature of the 3D Ising model. When $T>T_c$, the system energy $E$ and the average magnetization $\left|M\right|$ vary slowly with $T$. After $T$ goes below $T_c \approx 4.7J$, the energy $E$ decreases and $\left|M\right|$ increases (see Fig. \ref{fig_5}b). At $T=2.5J$, $\left|M\right|$ is saturated at the value 0.8, potentially due to the ineffectiveness of the Monte Carlo annealing. At a temperature $T=0.5J$, we obtain $\left|M\right|=0.93$ (see Fig. \ref{fig_5}c for ground state spin configuration). Due to the finite-size effect \cite{fisher1972scaling,landau1976finite}, the average magnetization $\left|M\right|$ changes slowly across $T=T_c$ (see Supplemental Materials). These results are consistent with previous numerical studies \cite{preis2009gpu,ghaemi2001calculation}, which demonstrates the generality of Fourier mask PIM in optical simulation of statistical physics. \par

\section{Discussions}
The SGMs have rugged energy landscape with many local minima \cite{binder1986spin}. The accurate evaluation of the Hamiltonian by the PIM allows an effective Monte Carlo annealing algorithm, significantly avoiding entrapment in local minima. Fourier-mask PIM can adapt to other lattice structures such as triangular, honeycomb and higher-dimensional lattices by arranging the configurations of spin sites on the SLM and designing the corresponding Fourier masks. The lattice size can be extended by increasing the active area on the SLM. {We can also design Fourier masks to simulate the SGMs with vacancy defects as well as in an external magnetic field.}  

\par

In the current Fourier-mask PIM, the speed is limited by the response time of the SLM, the frame rate of the CCD, and the computation of integration of $I(\boldsymbol{x})$ and $I_M$ on a computer. The latter can be improved by using optical meta-surfaces to work as a Fourier mask and directly obtain the Hadamard product from the light intensity \cite{liu2021multifunctional}. The CCD can be replaced by photodiodes to achieve summation without computers, which can further increases the sampling rate. We can also use the digital micromirror device to speed up the spin flip.\par

In conclusions, we develop a programmable Fourier-mask optical simulator for various SGMs. The Fourier mask can be obtained by making inverse Fourier transform of the couplings, and is additive for SGMs with complex short- and long-range couplings. Without sacrificing the speed, the Fourier-mask PIM can be used to simulate statistical models with a larger size, and can be implemented in multistate spins and high-dimensional lattices, providing a novel route to exploring the universality in phase transitions. The Fourier-mask PIM is also promising in combinatorial optimization problems of data learning theory \cite{watkin1993statistical}, error calibration \cite{sourlas1989spin}, and social group investigation \cite{korbel2023homophily}.\par

\section*{Methods}  
\subsection*{Experimental setup}\label{sec1}
A laser beam from a He-Ne laser (LASOS, LGK 7634) is expanded by two lenses (Lens1 and Lens2 with focal lengths of 50mm and 150mm), and is then shaped by a spatial filter, including an objective Obj1 (Olympus, 10×, NA=0.3), a pinhole (20 $\mathrm{\mu m}$), and Obj2 (Olympus, 4×, NA=0.13), in order to obtain a quasi-plane wave beam with 8 mm diameter. The laser beam then passes through a beam splitter (BS) and is modulated by a phase-only SLM (Holoeye, PLUTO-2.1, HED-6010-NIR-134). The modulated beam is Fourier transformed by Lens3 (Edmund, 32-327, achromat, focal length 100mm) and recorded by a CCD (AVT, Prosilica GT2460) at the back focal plane.

\subsection*{Designing the Fourier masks}
The Fourier mask is obtained from the inverse Fourier transform of the spin coupling $C_{j h}$. For NN couplings, $I_M$ is the summation of two cosine functions in $x$ and $y$ directions, $\mathrm{cos}(bx) +\mathrm{cos}(by)$, with values ranging from $-2$ to 2. In order to use optical elements to realize $I_M$, we shift and renormalize $I_M$ to values between 0 and 1. This can facilitate the future implementation of Fourier masks with a metasurface or an intensity attenuator. For the NN couplings, $I_M(\boldsymbol{x})$ is renormalized to $[\mathrm{cos}(bx) +\mathrm{cos}(by)+2]/{4}$. For the NNN couplings, the $I_M$ is renormalized to $[\mathrm{cos}(bx) \cdot\mathrm{cos}(by)+1]/{2}$. This procedure of renormalization introduces non-zero self spin couplings $C_{jj}={1}/{2}$, which can be eliminated by the unbiased normalization method. The Fourier mask for arbitrary-range couplings can be synthesised with the same procedure.

\subsection*{Calculating the Hamiltonian}
In order to obtain consistent results for different
illumination, we use an unbiased normalization approach to obtain the interaction Hamiltonian. First, all spins are set in the same direction and the corresponding intensity distribution $I_{\text{init}}(\boldsymbol{x})$ is obtained. The Hamiltonian for this particular spin configuration, denoted as $\mathcal{H}_{\text{init}}(\boldsymbol{x})$, is numerically calculated (e.g., for the 2D Ising model with NN couplings, it is $-2(N-\sqrt{N})$). {The self spin coupling term in Eq. (\ref{equ_2}) can be obtained by choosing a spin configuration with zero Hamiltonian, such as a stripe pattern for NN interaction. The intensity distribution of this configuration is recorded as $I_{\text{cali}}(\boldsymbol{x})$.} The Hamiltonian for an arbitrary spin configuration is
\begin{equation}
\mathcal{H}=\frac{ \mathcal{H}_{\text {init }}}{\mathcal{I}_{\text{init}}-\mathcal{I}_{\text{cali}}} \left(\int I_M(\boldsymbol{x}) I(\boldsymbol{x}) d \boldsymbol{x}-\mathcal{I}_{\text{cali}}\right),
\label{equ_s7}
\end{equation}
{where $\mathcal{I}_{\text{init}}=\int I_M(\boldsymbol{x}) I_{\text {init }}(\boldsymbol{x}) d \boldsymbol{x}$ and $\mathcal{I}_{\text{cali}}=\int I_M(\boldsymbol{x}) I_{\text {cali }}(\boldsymbol{x}) d \boldsymbol{x}$.} 

\subsection*{Sampling in simulated annealing}
The Hamiltonian obtained from the Fourier-mask PIM allows for the effective Metropolis-Hastings single-spin-flip algorithm during Monte Carlo annealing \cite{creutz1983monte}. In the experiment, the PIM is randomly initialized at a high temperature. We then perform single spin flips until the Markov chain reaches a stable state. The flip is accepted with a probability determined by the Metropolis-Hastings rule \cite{muller2023fast}, $P_{\text{acc}}=\min(1,e^{-\beta \Delta \mathcal{H}})$, where $\beta$ is the inverse temperature, given by $1/k_BT$ and the $\Delta \mathcal{H}=\mathcal{H}_2-\mathcal{H}_1$ is the energy difference between the spin configurations before ($\mathcal{H}_1$) and after ($\mathcal{H}_2$) one spin flipping. The spin configurations in this process form a Markov chain and eventually evolve into a stable state. At each effective temperature $T$, we obtain about 1,000 samples which satisfy the Boltzmann distribution. Notably, these samples are obtained from the Markov chain at fixed intervals, ensuring that they are independent and uniformly distributed for effective ensemble estimation \cite{yeomans1992statistical}.\par 

\section*{Data availability}
All data are available from the corresponding author on reasonable request.

\section*{Code availability}
All codes used to produce the findings of this study are available from the corresponding authors on reasonable request.

\section*{Acknowledgements}  
We thank Zhi-Chao Ruan for helpful discussion. This work was supported by the National
Natural Science Foundation of China (Grant No. 11934011),
National Key Research and Development Program of China (Grants No. 2019YFA0308100), the Strategic Priority
Research Program of Chinese Academy of Sciences (Grant
No. XDB28000000), and the Fundamental Research Funds
for the Central Universities.

\section*{Author contributions}
W.F. and D.W. conceived the idea and designed the experiment. W.F. and Y.S. carried out the experiment, collected data and performed numerical simulation. Y.S. wrote the control program. W.F., Y.S. and D.W. analyzed data and wrote the manuscript. All authors discussed the results and commented on the manuscript.

\section*{Competing interests}
The other authors declare no competing interests.

\end{document}


\title[Supplemental Materials: Programmable Photonic Simulator for Spin Glass Models]{Supplemental Materials: Programmable Photonic Simulator for Spin Glass Models}
\author*[1]{\fnm{Weiru} \sur{Fan}}\email{weiru\_fan@zju.edu.cn}
\equalcont{These authors contributed equally to this work.}

\author[1]{\fnm{Yuxuan} \sur{Sun}}\email{yuxuan\_sun@zju.edu.cn}
\equalcont{These authors contributed equally to this work.}

\author[1]{\fnm{Xingqi} \sur{Xu}}\email{xuxingqi@zju.edu.cn}

\author*[1,2,3,4]{\fnm{Da-Wei} \sur{Wang}}\email{dwwang@zju.edu.cn}

\author[1,2,3]{\fnm{Shi-Yao} \sur{Zhu}}\email{syzhu@zju.edu.cn}

\author[1]{\fnm{Hai-Qing} \sur{Lin}}\email{hqlin@zju.edu.cn}

\affil[1]{\orgdiv{Zhejiang Province Key Laboratory of Quantum Technology and Device, School of Physics, and State Key Laboratory for Extreme Photonics and Instrumentation}, \orgname{Zhejiang University}, \orgaddress{\city{Hangzhou}, \postcode{310027}, \state{Zhejiang Province}, \country{China}}}

\affil[2]{\orgdiv{College of Optical Science and Engineering}, \orgname{Zhejiang University}, \orgaddress{\city{Hangzhou}, \postcode{310027}, \state{Zhejiang Province}, \country{China}}}

\affil[3]{\orgname{Hefei National Laboratory}, \orgaddress{\city{Hefei}, \postcode{230088}, \country{China}}}

\affil[4]{\orgdiv{CAS Center for Excellence in Topological Quantum Computation}, \orgname{University of Chinese Academy of Sciences}, \orgaddress{\city{Beijing}, \postcode{100190}, \country{China}}}

\maketitle

\section{Experiment Setup}\label{sec1}

\begin{figure}[H]%
\centering
\includegraphics[width=0.9\textwidth]{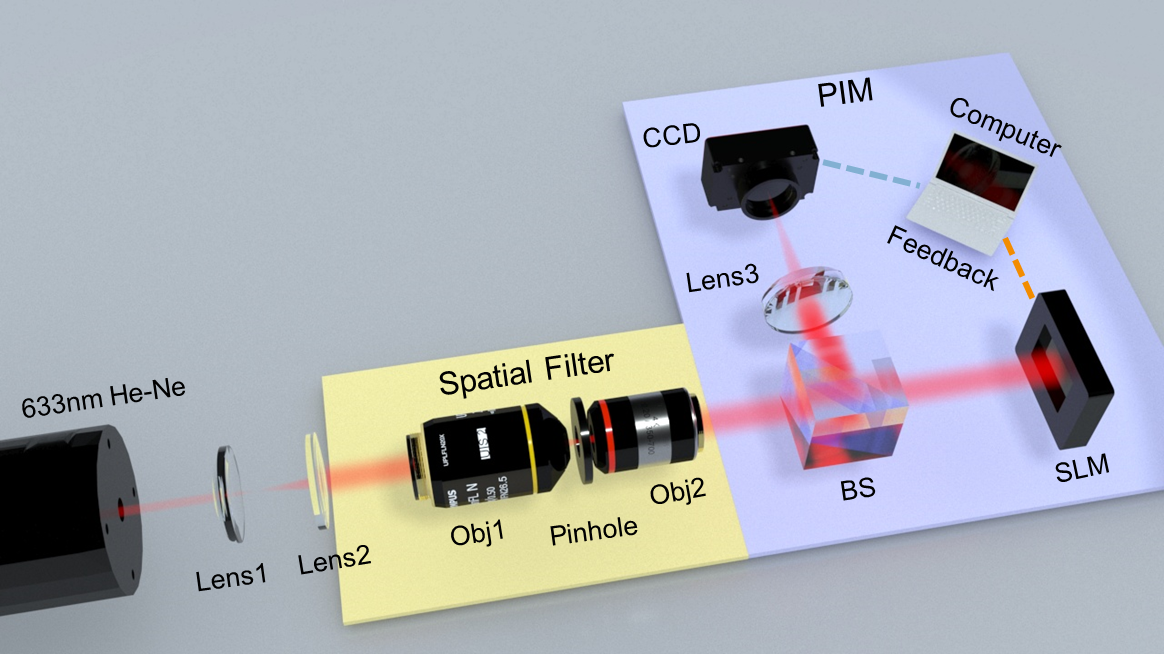}
\caption{The experimental setup of Fourier-mask PIM. A laser beam from a He-Ne laser (LASOS, LGK 7634) is expanded by Lens1 and Lens2 with focal lengths of 50mm and 150mm. The beam is then shaped by a spatial filter, which includes Obj1 (Olympus, 10×, NA=0.3), a pinhole (20 $\mathrm{\mu m}$), and Obj2 (Olympus, 4×, NA=0.13), in order to achieve a quasi-plane wave with 8 mm diameter. The laser beam then passes through a beam splitter (BS) and is modulated by a phase-only SLM (Holoeye, PLUTO-2.1, HED-6010-NIR-134). The modulated beam is transformed by Lens3 (Edmund, 32-327, achromat with a focal length of 100mm) to perform the Fourier transform and is recorded by a CCD (AVT, Prosilica GT2460) at the back focal plane.}\label{fig1}
\end{figure}\par

\section{The Fourier Mask}
With the gauge transformation method \cite{fang2021experimental} each effective spin is encoded on an individual area on the SLM. The modulated light field is,
\begin{equation}
E(\boldsymbol{u})=E_0 \sum_{j=1}^N \sigma_j e^{i\arccos \xi_j} \operatorname{rect}\left(\frac{\boldsymbol{u}-\boldsymbol{u}_j}{W}\right),
\label{equ_s1}
\end{equation}
where $E_0$ is the amplitude of the light, $\boldsymbol{u}$ are the spatial coordinates on the SLM, $\boldsymbol{u}_j$ is the position of the $j$th pixel, and $W$ is the width of a pixel. The rectangular function is defined as,
\begin{equation}
\operatorname{rect}\left(\frac{\boldsymbol{x}}{W}\right)= \begin{cases}1 & |x|,|y| \leq W / 2 \\ 0 & |x|,|y|>W / 2\end{cases}.
\label{equ_s2}
\end{equation}
After passing through a lens, the light intensity $I(\boldsymbol{x})$ on the Fourier plane is \cite{goodman2005introduction},
\begin{equation}
I(\boldsymbol{x})=I_0 \sum_{j, h=1}^N \xi_j \xi_h \sigma_j \sigma_h \exp \left(i \frac{2 \pi}{f \lambda}\left(\boldsymbol{u}_j-\boldsymbol{u}_h\right) \cdot \boldsymbol{x}\right) \operatorname{sinc}^2\left(\frac{\boldsymbol{x} W}{f \lambda}\right),
\label{equ_s3}
\end{equation}
where $I_0$ is the intensity factor, $f$ is the focal length, $\lambda$ is the wavelength, and $\boldsymbol{x}$ is the coordinate on the Fourier plane. The intensity is recorded by a CCD and we make integration of it with the Fourier mask $I_M(\boldsymbol{x})$ (the Hadamard product and summation),
 \begin{equation}
\begin{aligned}
& \int I(\boldsymbol{x}) I_M(\boldsymbol{x}) d \boldsymbol{x} \\
= & I_0 \sum_{j, h=1}^N \xi_j \xi_h \sigma_j \sigma_h \int I_M(\boldsymbol{x}) \exp \left(i \frac{2 \pi}{f \lambda}\left(\boldsymbol{u}_j-\boldsymbol{u}_h\right) \cdot \boldsymbol{x}\right) \operatorname{sinc}^2\left(\frac{\boldsymbol{x} W}{f \lambda}\right) d \boldsymbol{x} \\
= & I_0 \sum_{j,h=1}^N C_{j h} \xi_j \xi_h \sigma_j \sigma_h\\
= & I_0 \sum_{j\neq h=1}^N C_{j h} \xi_j \xi_h \sigma_j \sigma_h+cI_0N,
\end{aligned}
\label{equ_s4}
\end{equation}
where $C_{jh}=\int I_M(\boldsymbol{x}) \exp \left(i \frac{2 \pi}{f \lambda}\left(\boldsymbol{u}_j-
\boldsymbol{u}_h\right) \cdot \boldsymbol{x}\right) \operatorname{sinc}^2\left(\frac{\boldsymbol{x} W}{f \lambda}\right) d \boldsymbol{x}$ is the spin coupling, {and the self coupling $C_{jj}$ is defined as $c$ here and can be eliminated by the unbiased normalization method described in Sec. \ref{sec_S3}.} From Eq. (\ref{equ_s4}) we can design an arbitrary Hamiltonian of the SGM from a proper Fourier mask. With negligible size effect of $W$, i.e., $\operatorname{sinc}^2\left(\frac{\boldsymbol{x} W}{f \lambda}\right)\sim 1$, the Fourier mask $I_M(\boldsymbol{x})$ can be considered as the inverse Fourier transform $(\mathcal{F}^{-1})$ of the spin coupling $C_{jh}$,
\begin{equation}
I_M(\boldsymbol{x})=\frac{1}{2 \pi} \int C_{j 1} \exp \left(-i \frac{2 \pi}{f \lambda}\boldsymbol{u}_j\cdot \boldsymbol{x}\right) d\boldsymbol{u}_j=\mathcal{F}^{-1}\left(C_{j1}\right).
\label{equ_s5}
\end{equation}
{Since $C_{jh}$ is only related to the relative positions of the two spins ($\boldsymbol{u}_j-\boldsymbol{u}_h$) and the lattice has translational symmetry, we set $h=1$ and define $\boldsymbol{u}_1=0$.}\par 

{The Fourier mask $I_M$ is the summation of two cosine functions in $x$ and $y$ directions $\mathrm{cos}(bx) +\mathrm{cos}(by)$ for the NN couplings, with values ranging from -2 to 2. In order to use optical elements to realize $I_M$, We shift and renormalize $I_M$ to values between 0 and 1. This can facilitate the future implementation of Fourier mask with a metasurface or an intensity attenuator. For the NN couplings, $I_M(\boldsymbol{x})$ is renormalized to $[\mathrm{cos}(bx) +\mathrm{cos}(by)+2]/{4}$. For the NNN couplings, the $I_M$ is renormalized to $[\mathrm{cos}(bx) \cdot\mathrm{cos}(by)+1]/{2}$. This procedure of renormalization introduces non-zero self spin couplings $C_{jj}={1}/{2}$, which can be eliminated by the unbiased normalization method in the next section.}

\section{Unbiased Normalization}
\label{sec_S3}
The integral in Eq. (\ref{equ_s4}) is different from the Hamiltonian in Eq. (1) by an intensity factor $I_0$, which needs to be removed for consistent results under different illumination powers. In the experiment, the Gaussian light beam is focused on the Fourier plane, which generates an Airy disk. There are two sources of error that we need to consider. First, the finite size of the Airy disk results in difficulty in obtaining the intensity factor $I_0$. Second, $I(\boldsymbol{x})$ and $I_M(\boldsymbol{x})$ can have a position shift in experiment, which results in inaccuracy in the evaluation of the Hamiltonian. We first evaluate the error due to a position shift $\boldsymbol{x}_0$, 
\begin{equation}
\begin{aligned}
& \int I(\boldsymbol{x}) I_M\left(\boldsymbol{x}-\boldsymbol{x}_0\right) d \boldsymbol{x} \\
= & I_0 \sum_{j, h=1}^N \xi_j \xi_h \sigma_j \sigma_h \exp \left(i \frac{2 \pi}{f \lambda}\left(\boldsymbol{u}_j-\boldsymbol{u}_h\right) \cdot \boldsymbol{x}_0\right) \\
&\cdot\int I_M(\boldsymbol{x}) \exp \left(i \frac{2 \pi}{f \lambda}\left(\boldsymbol{u}_j-\boldsymbol{u}_h\right) \cdot \boldsymbol{x}\right) \operatorname{sinc}^2\left(\frac{\boldsymbol{x} W}{f \lambda}\right) d \boldsymbol{x} \\
= & I_0 \sum_{j, h=1}^N \cos \left(\frac{2 \pi}{f \lambda}\left(\boldsymbol{u}_j-\boldsymbol{u}_h\right) \cdot \boldsymbol{x}_0\right) C_{j h} \xi_j \xi_h \sigma_j \sigma_h.
\end{aligned}
\label{equ_s6}
\end{equation}
In our experiment, $\boldsymbol{x}_0$ is small enough such that the cosine fucntion is approximately one. Therefore, the main deviation comes from the finite size effect of the Airy disk in determining $I_0$.\par

In order to address this issue, we introduce an unbiased normalization approach to obtain the Hamiltonian. We first set all spins in the same direction and obtain the corresponding intensity distribution $I_{\text{init}}(\boldsymbol{x})$. The Hamiltonian for this particular spin configuration, denoted as $\mathcal{H}_{\text{init}}(\boldsymbol{x})$, is numerically calculated (e.g., for the 2D Ising model with NN couplings, it is $-2(N-\sqrt{N})$). {The self spin couplings term in Eq. (\ref{equ_s4}) can be obtained by choosing a spin configuration with zero Hamiltonian, such as a stripe pattern for NN interaction. The intensity distribution of this configuration is recorded as $I_{\text{cali}}(\boldsymbol{x})$.} The Hamiltonian for an arbitrary spin configuration is
\begin{equation}
\mathcal{H}=\frac{\mathcal{H}_{\text {init }}}{\mathcal{I}_{\text{init}}-\mathcal{I}_{\text{cali}}} \left(\int I_M(\boldsymbol{x}) I(\boldsymbol{x}) d \boldsymbol{x}-\mathcal{I}_{\text{cali}}\right),
\label{equ_s7}
\end{equation}
{where $\mathcal{I}_{\text{init}}=\int I_M(\boldsymbol{x}) I_{\text {init }}(\boldsymbol{x}) d \boldsymbol{x}$ and $\mathcal{I}_{\text{cali}}=\int I_M(\boldsymbol{x}) I_{\text {cali }}(\boldsymbol{x}) d \boldsymbol{x}$.} 


\begin{figure}[H]%
\centering
\includegraphics[width=1\textwidth]{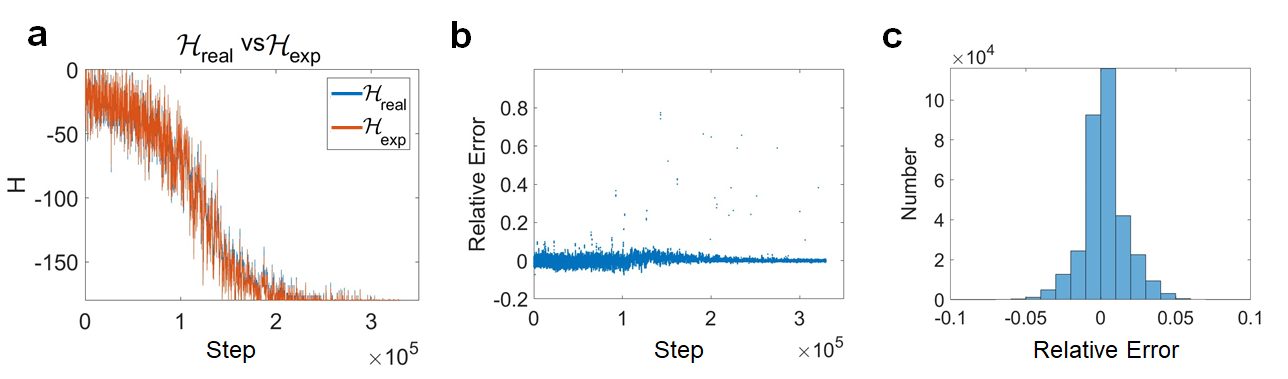}
\caption{The unbiased normalization. (a) The evolution of the Hamiltonian of the Mattis SGM with NN coupling and $N=100$, during the Monte Carlo annealing; (b) The relative error obtained by $\hat{\boldsymbol{e}}=(\mathcal{H}_{\text{exp}}-\mathcal{H}_{\text{real}})/{\mathcal{H}_{\text{init}}}$; (c) The histogram of the relative error. $\mathcal{H}_{\text{exp}}$ is the Hamiltonian obtained from the unbiased normalization, while the $\mathcal{H}_{\text{real}}$ is the Hamiltonian numerically computed from Eq. (1).}\label{fig2}
\end{figure}

\section{Monte Carlo Annealing Algorithm}
The Hamiltonian obtained from the Fourier-mask PIM allows for the effective Metropolis-Hastings single-spin-flip algorithm during Monte Carlo annealing \cite{creutz1983monte}. In this framework, we start from a random spin configuration $\left\{\sigma_1\right\}$, and the corresponding Hamiltonian $\mathcal{H}_1$ is obtained from the PIM. A lattice site is randomly selected, and its spin is flipped, resulting in a new spin configuration $\left\{\sigma_2\right\}$. The corresponding Hamiltonian $\mathcal{H}_2$ is obtained from the PIM. The energy difference $\Delta\mathcal{H}=\mathcal{H}_2-\mathcal{H}_1$ is subsequently computed as the criterion for acceptance. If $\Delta\mathcal{H}$ is negative, indicating a decrease in energy after the spin flip, the flip is accepted, and we continue the spin flip procedure by replacing $\left\{\sigma_1\right\}$ with $\left\{\sigma_2\right\}$. On the other hand, if the energy difference $\Delta\mathcal{H}$ is positive, the flip is accepted with a probability ($P_{\text{acc}}$) determined by the Metropolis-Hastings rule,
\begin{equation}
P_{\text{acc}}=\min\left\{1,~\exp (-\beta \Delta \mathcal{H})\right\}
\label{equ_s8}
\end{equation}
where $\beta$ is the inverse temperature, given by $1/k_BT$. The spin configurations in this process form a Markov chain and eventually evolve into a stable state. 

In the experiment, the PIM is randomly initialized at a high temperature, and performs single spin flips until the Markov chain reaches the stable state. At each effective temperature $T$, we obtain $L$ samples from the PIM, denoted as $\left\{\left\{\sigma_{T,1}\right\},\left\{\sigma_{T,2}\right\},\cdots,\left\{\sigma_{T,L}\right\}\right\}$, which satisfy the Boltzmann distribution. Notably, these samples are obtained by sampling the Markov chain at fixed intervals, ensuring that they are independent and uniformly distributed for effective ensemble estimation.\par 

Using these samples, various physical quantities, such as the average magnetization $M$, magnetic susceptibility $\chi$, and heat capacity $C_V$, can be evaluated by averaging over the $L$ samples \cite{yeomans1992statistical}. We repeat the experimental procedure at different temperatures. By examining the behavior of these physical quantities, we examine the phase transition of the system.\par 

With the gauge transformation, the Monte Carlo evolution of the effective spins $\left\{\sigma_j^{\prime}\right\}$ are equivalent to the evolution of the original spins $\left\{\sigma_j\right\}$ with arbitrary coupling configurations $\left\{\xi_j\right\}$. Consequently, only one Monte Carlo evolution of the effective spins $\left\{\sigma_j^{\prime}\right\}$ is needed. The original spins $\left\{\sigma_j\right\}$ with a specific coupling configuration $\left\{\xi_j\right\}$ can be obtained from the effective spins $\left\{\sigma_j^{\prime}\right\}$ by simply applying the inverse gauge transformation $\sigma_j=\sigma_j^{\prime}/\xi_j$. This approach allows for a more efficient Monte Carlo computation.

\section{The Fourier Mask for 3D Ising Model with NN Coupling}
\begin{figure}[H]%
\centering
\includegraphics[width=0.8\textwidth]{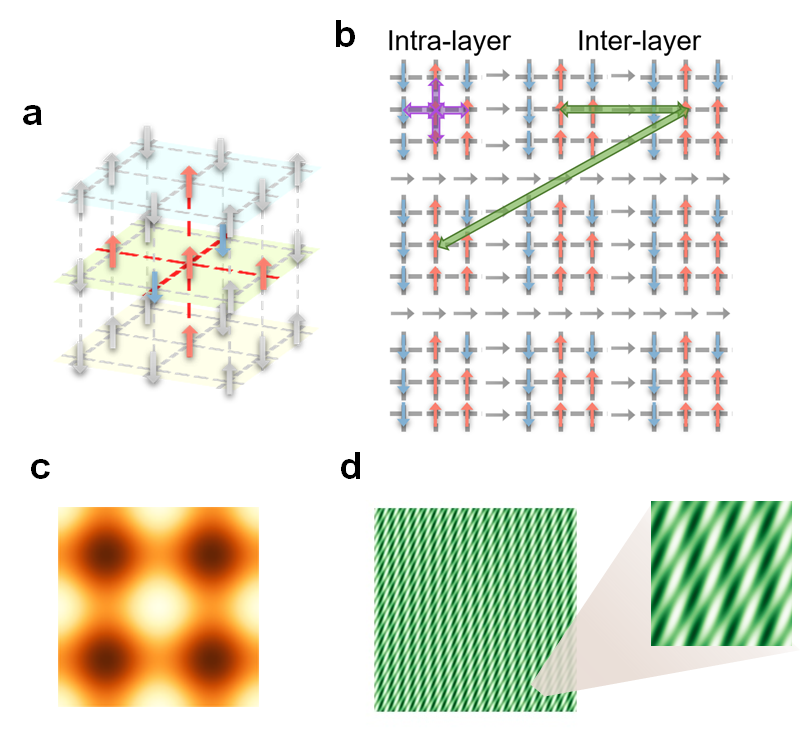}
\caption{The 3D Ising model and its Fourier mask. (a) A schematic of the Ising model on a 3D simple cubic lattice with NN coupling. (b) The interactions arranged in a 2D plane for constructing the 3D Ising model. (c) and (d) are the Fourier masks corresponding to the intra-layer and inter-layer couplings. The purple arrows represent the intra-layer NN coupling, while the green arrows represent the inter-layer coupling. The 9 layers are arranged on the 2D SLM plane as a 3×3 array. Each layer consists of 9×9 spins. There is a one-pixel gap between every layer, which are not modulated to prevent unwanted inter-layer couplings, as shown by the gray arrows.}\label{fig3}
\end{figure}

To realize the 3D Ising model in the 2D PIM, we need both long-range and short-range interactions in the 2D lattice to obtain the corresponding inter-layer and intra-layer couplings. In this strategy, the two types of couplings can be expressed as follows:
\begin{equation}
\begin{aligned}
 C_{j h, \text { intra }}=&\delta\left(u_j-u_h\right) \delta\left(v_j-v_h-s\right)+\delta\left(u_j-u_h\right) \delta\left(v_j-v_h+s\right) \\
& +\delta\left(u_j-u_h-s\right) \delta\left(v_j-v_h\right)+\delta\left(u_j-u_h+s\right) \delta\left(v_j-v_h\right),
 \end{aligned}
 \label{equ_s9}
 \end{equation}
\begin{equation}
\begin{aligned}
 C_{j h, \text { inter }}=&\delta\left(u_j-u_h-9 s\right) \delta\left(v_j-v_h\right)+\delta\left(u_j-u_h+9 s\right) \delta\left(v_j-v_h\right) \\
& +\delta\left(u_j-u_h-18 s\right) \delta\left(v_j-v_h-9 s\right)+\delta\left(u_j-u_h+18 s\right) \delta\left(v_j-v_h+9 s\right),
\end{aligned}
 \label{equ_s10}
\end{equation}
where $s$ is the lattice constant, and $\boldsymbol{u}_j = (u_j,v_j)$ is the spatial coordinates of the 2D lattice. Therefore, the Fourier mask is designed as,
\begin{equation}
I_{M,\text{intra}}(\boldsymbol{x})=\operatorname{cos}bx+\operatorname{cos}by,
\label{equ_s11}
\end{equation}
\begin{equation}
I_{M,\text{inter}}(\boldsymbol{x})=\operatorname{cos}9bx+\operatorname{cos}\left(18bx+9by\right).
\label{equ_s12}
\end{equation}
By substituting Eq. (\ref{equ_s11}) and Eq. (\ref{equ_s12}) into Eq. (\ref{equ_s4}), the value of $b$ can be calculated (see the two Fourier masks are in Fig. \ref{fig3}c and \ref{fig3}d). These two Fourier masks are added to form a single Fourier mask for the 3D Ising model. This approch can be used for higher-dimensional lattices.

\begin{figure}[H]%
\centering
\includegraphics[width=1\textwidth]{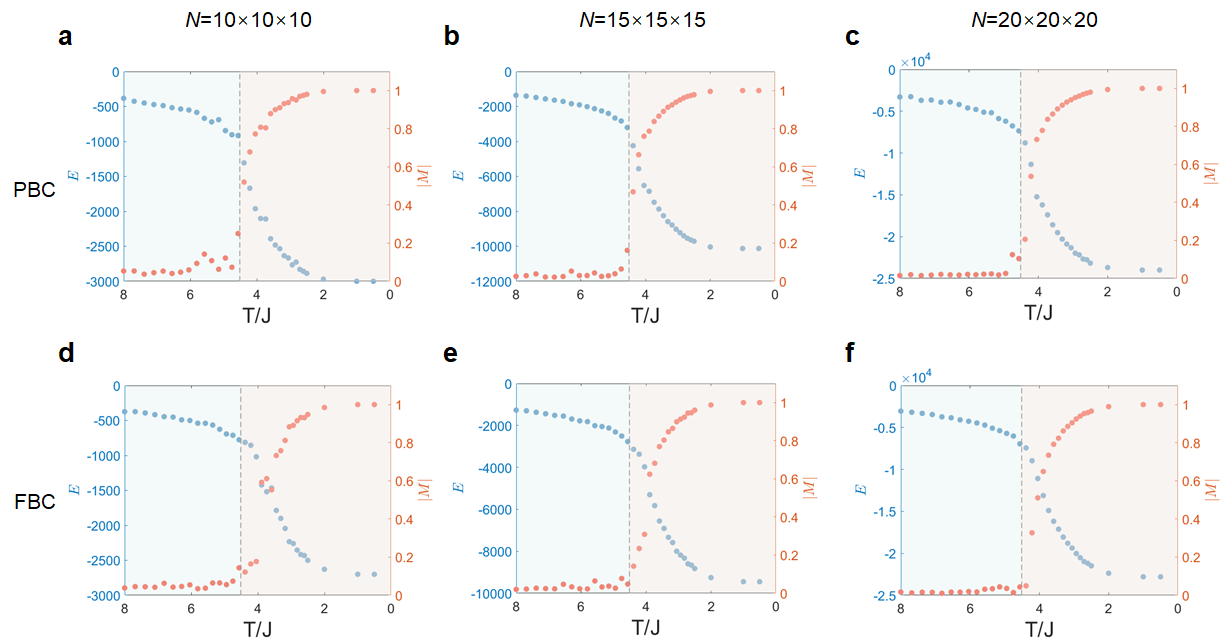}
\caption{The simulation of 3D Ising model phase transition with different sizes and boundary conditions. (a)-(c) The values of average magnetic moment $\left|M\right|$ and average energy $E$ as functions of temperature $T$ for 3D Ising model with spin sizes $N=10\times10\times10$, $N=15\times15\times15$, and $N=20\times20\times20$, under PBC. (d)-(f) 
have the corresponding parameters of (a)-(c) but with FBC.}
\label{fig4}
\end{figure}

The absence of a sharp change of the order parameter $\left|M\right|$ at the critical temperature in main text FIG.~5 is attributed to the finite-size effect \cite{cardy2012finite}. For an infinite 3D lattice, the correlation length $\zeta$ diverges at $T=T_c$. However, the lattice size is always finite in simulation, such as $l\times l\times l$, where $l$ is the lattice length in each dimension. Therefore, $\zeta$ is limited by the lattice size, which results in size-dependent thermodynamic quantities. For a finite lattice with periodic boundary condition (PBC) when $T$ increases above $T_c$, $\left|M\right|$ reduces to zero in a slower manner for smaller $l$ (Fig. \ref{fig4}a). However, for $T<T_c$ the change in $\left|M\right|$ is almost independent of the lattice size. In contrast, with the free boundary condition (FBC) the change in $\left|M\right|$ strongly depends on lattice size for $T<T_c$ (Fig. \ref{fig4}d-f) \cite{landau1976finite}, which can explain the absence of a sharp change of $\left|M\right|$ in the experimental results in FIG.~5.